\documentclass[twocolumn,showpacs,preprintnumbers,amsmath,amssymb]{revtex4}

\usepackage{graphicx}
\usepackage{dcolumn}
\usepackage{bm}
\usepackage{amsmath,dsfont}
\usepackage{amssymb,amsthm,amscd, amsbsy, array}

\newcommand{\half}{{\scriptstyle{\frac{1}{2}}}}
\def\2{{\half}}

\def\dAlembert{\vcenter {
    \hbox {\vrule height8pt width0.4pt depth0.0pt
           \vrule height8pt width7.2pt depth-7.6pt
           \vrule height8pt width0.4pt depth0.0pt
           \kern-8pt
           \vrule height0.4pt width8pt depth0.0pt
          \,}}}

\def\beq{\begin{equation}}
\def\eeq{\end{equation}}
\def\beqa{\begin{eqnarray}}
\def\eeqa{\end{eqnarray}}
\def\nn{\nonumber}

\def\tr{{\,\rm tr\,}}
\def\II{{\mathds{1}}}
\def\smallover#1/#2{\hbox{$\textstyle\frac{#1}{#2}$}} 

\begin{document}


\title{Bogomolny equations and conformal
transformations in curved space}

\author{P-M Zhang}\email{zhpm-at-impcas.ac.cn}
\affiliation{Institute of Modern Physics, Chinese Academy of Sciences
\\
Lanzhou (China)
}

\author{P.~A.~Horv\'athy}\email{horvathy-at-lmpt.univ-tours.fr}
\affiliation{Laboratoire de Math\'ematiques et de Physique
Th\'eorique\\
 Universit\'e de Tours 
(France)}

\date{\today}

\begin{abstract}{\it The coupling of the Higgs field
through the Ricci tensor, put forward  by Balakrishna and Wali, is derived using a conformal rescaling of the metric. Earlier results on ``Bogomolny-type'' equations in curved space, by Comtet, and others,
are recovered. The procedure can be generalized to
any static background metric.
}
\end{abstract}


\maketitle


\section{Introduction}

In the study of solitons in flat space,
the `Bogomolny', or `self-duality' equations \cite{Bogom},
$ 
D_i\Phi=\2\epsilon_{ijk}F^{jk},
$ 
 play two, complementary roles. On the one hand, they
 provide the absolute minima of the energy, yielding static solutions.  But they  also 
 allow us to reduce the second-order 
 field equations to first-order ones.

In curved space, an obstruction arises, though~: a
self-dual field may fail to solve the field equations.

To be specific,
consider  a purely magnetic, static Yang-Mills-Higgs system $(A_i,\phi)$ in $3$ space dimensions,
where the YM potential $A_i$ takes its values in the Lie algebra of the
gauge group  $G$ (a compact Lie group),
$F_{ij}=\partial_iA_j-\partial_jA_i+[A_i,A_j]$
is the YM field strength and $\phi$,
the Higgs field, belongs to
the adjoint representation of the gauge group. Let
$g_{\mu\nu}=\big(g_{oo},-\hat{g}_{ij}\big)$
be a static  background metric.
For vanishing Higgs potential,
the Lagrangian is
\beq
L_{YMH}=\tr\Big(-\smallover1/4 F_{ij}F^{ij}+\2D_i\phi\, 
D^i\phi\Big)\sqrt{g}\,,
\label{1.1}
\eeq
where $g=g_{oo}\hat{g}$ is the determinant of metric 
with $\hat{g}=\det(g_{ij})$ the determinant of the space-metric alone.
The associated field equations read
\beq\left\{
\begin{array}{lll}
\displaystyle\frac
{1}{\sqrt{g}}D_i\big(\sqrt{g}\,F^{ij}\big)
&=&[\phi,D^j\phi],
\\[8pt]
\displaystyle\frac
{1}{\sqrt{g}}D_i\big(\sqrt{g}\,D^i\phi\big)
&=&0.
\end{array}\right.
\label{1.2}
\eeq

The natural generalization to curved space of the Bogomolny equations
would be
\beq
D_i\phi=\2\sqrt{\hat{g}}\,\epsilon_{ijk}F^{jk}
\label{1.3}
\eeq
However,
inserting Eq. (\ref{1.3}) into the r.h.s. of the second equation
in (\ref{1.2}) yields,  using the Bianchi identities,
$
\2\partial^i\big(\ln g_{oo}\big)D_i\phi,
$
which does not vanish, unless $g_{oo}$ is a constant.

A clever way of removing this
obstruction is to add a suitable curvature
term to the Lagrangian  \cite{BW,Edel}. 
One assumes  that the metric is of the 
 Papapetrou-Majumdar form 
\beq
V^2dt^2-\frac{d\vec{x}^{2}}{V^2},
\qquad
\label{2.4-5}
\eeq
and one seeks  static fields
$\phi, A_i$ which extremize  
\beq
S=\int\tr\left[-\frac{R}{4}\,\psi^2
-\frac{1}{4}\big(F_{ij}F^{ij}\big)
+\frac{1}{2}\big(D_i\psi D^i\psi\big)
\right]\,V^{-2}d^4{x},
\label{3.1}\eeq
where $R$ is the Ricci scalar of the background metric. 
This  expression differs from (\ref{1.1}) in the term
$(R/4)\tr\psi^2$, which makes it indefinite.
The associated  field equations,
\beq\left\{
\begin{array}{lll}
D_i(V^2F_{ij})&=&[\psi, D_j\psi],
\\[8pt]
V^2D_i^2\psi&=&\frac{1}{2}R\,\psi.
\end{array}\right.
\label{3.2}
\eeq
can be obtained by solving instead \footnote{ 
The Einstein equations can also  be solved when
$
\tr\,\psi^2\equiv1
$ 
everywhere in space \cite{BW,FHH}.},
\beq
F_{ij}=\pm\frac{1}{V^2}\epsilon_{ijk}D^k(V\psi).
\label{3.3} 
\eeq

To explain why this works, we note first that (\ref{3.3}) are in fact  (\ref{1.3}) for the conformally rescaled metric
$
G_{\mu\nu}=V^{-2}\,g_{\mu\nu}
=\big(1,-V^{-4}\II_3\big),
$
 for which no obstruction arises. But 
such a rescaling, implemented on the fields $(A_i,\phi)$ as
$
A_i\to A_i,
\,
\phi\to\psi=V^{-1}\phi,
$ 
changes (\ref{1.1}) written in the rescaled metric $G_{ij}$ into
\beqa
&\tr\left[
-\smallover1/4 F_{ij}F^{ij}
+\2D_i\psi D^i\psi\right]\sqrt{-g}
+\partial_j\left[\2\sqrt{g}\,\partial^i\ln\sqrt{g_{00}}\tr\psi^2\right]
\nn
\\[6pt]
&-\2\sqrt{g}\tr\psi^2\widehat{\bigtriangleup}\ln\sqrt{g_{00}},
\label{3.7}
\eeqa
where $\widehat{\bigtriangleup}=\frac{1}{\sqrt{\hat{g}}}\,
\partial_i\big(\sqrt{\hat{g}}\,\partial^i\big)$ is the Laplacian associated
with the space metric $\hat{g}_{ij}$.
Our clue is now that, for the class (\ref{2.4-5}) of metrics,
\beq
\widehat{\bigtriangleup}\ln\sqrt{g_{00}}=\frac{1}{2}R,
\label{3.8}
\eeq
 one half of the Ricci scalar.
In the last term in (\ref{3.7}) we recognize, hence, precisely the  curvature term  in (\ref{3.1}). 
The 
rescaled expression, (\ref{3.7}),  only 
differs from the  density (\ref{3.1})
by a surface
term; they are, therefore, equivalent.
At last, the self-duality equations are conformally invariant.

\goodbreak

Our results here shed some new light on those, obtained
earlier by Comtet,  and others \cite{CFH}. In that approach,
the original field equations, (\ref{1.2}), are kept unchanged but
the Bogomolny equations are modified. 
\beq
D_i\phi+\partial_i\ln\sqrt{g_{00}}\,\phi=\pm\2\sqrt{\hat{g}}\,\epsilon_{ijk}F^{jk}
\label{Bogotype}
\eeq
is readily shown to
solve the second-order field equations (\ref{1.2}), provided
the metric satisfies the constraint
\beq
\widehat{\bigtriangleup}\ln\sqrt{g_{00}}
=0.
\label{2.2}
\eeq
Indeed, if the constraint (\ref{2.2}) holds then the Ricci scalar vanishes by (\ref{3.8}),
so that the curvature-modified model (\ref{3.1}) 
reduces to (\ref{1.1}). Note that for (\ref{2.2})
$V$ takes the Papapetrou-Majumdar form
$V^{-1}=1+\sum_a^N\frac{m_a}{|\vec{r}-\vec{r}_a|}
$,
whose particular case is
the extreme Reissner-Nordstr\"om metric
$V^{-1}=1-1/r$ considered in \cite{CFH}.

On the other hand, the Bogomolny equation (\ref{1.3}) for the rescaled metric 
$G_{\mu\nu}=g_{\mu\nu}/g_{oo}$ becomes, for $\psi=\phi/\sqrt{g_{oo}}$, precisely
(\ref{Bogotype}).

Which one is the `natural' theory: that in
 (\ref{1.1}), or the curvature-modified
 one in (\ref{3.1})~? An argument
in favor of the second choice is the following:
a static and purely magnetic theory can be converted, by putting $A^a_0=\sqrt{g_{oo}}\,\psi^a$,
into a pure Yang-Mills configuration on euclidean four-space. Then essentially the same calculation as for 
(\ref{3.7}) shows that the curvature-modified action
(\ref{3.1}) [and \emph{not} the simple expression
(\ref{1.1})] becomes, up to a surface term, the
pure Yang-Mills expression
$-\smallover{1}/{4}
\tr F_{\mu\nu} F^{\mu\nu}\sqrt{g}$,
 whereas the $4D$ self-duality equations,
 $F_{\mu\nu}=\pm\2\sqrt{g}\epsilon_{\mu\nu\rho\sigma}F^{\rho\sigma}$, become (\ref{Bogotype}) \cite{CFH}.

 Our results have the following intuitive explanation:
 the self-duality equations are conformally invariant,
 but the field theory defined by (\ref{1.1}) are not.
 However, 
adding the curvature term restores  the conformal 
invariance, as it has been observed before in the Chern-Simons context \cite{DHP1}.

It is worth mentioning that the procedure considered here can be further generalized. Let indeed 
$g_{\mu\nu}=(g_{oo},-\hat{g}_{ij})$ 
be an arbitrary static spacetime, and 
consider the rescaled metric $G_{\mu\nu}=(1,-\widehat{G}_{ij})$, $\widehat{G}_{ij}=\hat{g}_{ij}/g_{oo}$.
Then  for the YMH theory  (\ref{1.1}) with the rescaled metric $G_{\mu\nu}$,
the Bogomolny (alias self-dual) equations
(\ref{1.3}) work. Then implementing the rescaling as 
$\psi=(g_{oo})^{-1/2}\phi$ provides us with a 
YMH theory on the original space with Lagrangian
\beqa
\tr\left[-\frac{1}{4}F_{ij}F^{ij}+\frac{1}{2}D_i\psi D^i\psi +\frac{1}{2}\Omega\,\psi^2\right]\sqrt{g},
\label{genrescac}
\\[10pt]
\Omega=-\widehat{\Delta}\ln\sqrt{g_{00}}
\label{Omega}
\eeqa
where a surface term has been dropped. The
Bogomolny equations (\ref{1.3}) become ``Bogomolny-type'' , Eqns. (\ref{Bogotype}).

In the Papapetrou-Majumdar case, $\Omega=-R/2$, and we
recover the results in \cite{BW}.

Another interesting case can be that of AdS space \footnote{Monopoles in AdS space were considered by
Lugo et al \cite{LMS}. The do not have Bogomolny equations, however, since their model is different.
In particular, they don't have our $\Omega$-term here.},
\begin{equation}
ds^2=\left( 1-\frac \Lambda 3r^2\right) dt^2-\frac{dr^2}{1-\frac \Lambda 3r^2%
}-r^2d\omega,
\label{AdSmetric}
\end{equation}
where $d\omega=d\theta^2+\sin^2\theta d\phi^2$. 
Then we find
\beq
\Omega=\frac{1-\frac{2}{9}\Lambda r^2}{1-\frac{1}{3}\Lambda r^2}\,\Lambda.
\label{AdSomega}
\eeq
For large $r$, (\ref{AdSomega}) becomes approximately
$\Omega\approx R/6$, where  $R=4\Lambda$ is the scalar
curvature.

Similarly in Schwarzschild space,
\begin{equation}
ds^2=\left(1-\frac{r_c}r\right)dt^2-\frac{1}{\left( 1-\dfrac{r_c}r\right)}dr^2-r^2d\omega,
\label{Schwarzschild}
\end{equation}
we have
\beq
\Omega=\frac{1}{4}\frac{r_c^2}{r^4(1-\dfrac{r_c}{r})}\,.
\eeq
In the Reissner-Nordstr\"om case, at last,
\begin{equation}
ds^2=(1-\frac{r_s}r+\frac{r_Q^2}{r^2})dt^2-\frac{dr^2}{(1-\dfrac{r_s}r+\dfrac{
r_Q^2}{r^2})}-r^2d\omega\,,
\end{equation}
we find
\beq
\Omega =\frac{1}{4}\frac{r_s^2-4r_Q^2}{r^4(1-\dfrac{r_s}r+\dfrac{r_Q^2}{r^2})},
\eeq
which, in the extreme case $r_s^2=4r_Q^2$, vanishes, 
leaving us with the result of \cite{CFH}.

\begin{acknowledgments}
P.A.H is indebted to the \textit{
Institute of Modern Physics} (Lanzhou) of
the Chinese Academy of Sciences for hospitality.
\end{acknowledgments}


\end{document}